\documentclass{article}

\usepackage{arxiv}

\usepackage[utf8]{inputenc} 
\usepackage[T1]{fontenc}    
\usepackage[numbers]{natbib}
\usepackage{hyperref}       
\usepackage{url}            
\usepackage{amsmath}
\usepackage{booktabs}       
\usepackage{amsfonts}       
\usepackage{nicefrac}       
\usepackage{microtype}      
\usepackage{lipsum}
\usepackage{graphicx}
\usepackage{multirow}
\usepackage{threeparttable}
\usepackage{algorithm}
\usepackage{algpseudocode}
\usepackage{balance}
\graphicspath{ {./images/} }

\providecommand{\Description}[1]{}

\title{Adaptive Defense Orchestration for RAG: A Sentinel-Strategist Architecture against Multi-Vector Attacks}

\author{
  Pranav Pallerla \\
  School of Computer and Information Sciences \\
  University of Hyderabad \\
  Hyderabad, Telangana, India \\
  \texttt{pallerlapranavdec27@gmail.com} \\
  \And
  Wilson Naik Bhukya \\
  School of Computer and Information Sciences \\
  University of Hyderabad \\
  Hyderabad, India \\
  \texttt{rathore@uohyd.ac.in} \\
  \And
  Bharath Vemula \\
  Purdue University \\
  West Lafayette, USA \\
  \texttt{vemula.sai8@gmail.com} 
  \And
  Charan Ramtej Kodi \\
  School of Computer and Information Sciences \\
  University of Hyderabad \\
  Hyderabad, India \\
  \texttt{21mcpc10@uohyd.ac.in} 
}

\begin{document}
\maketitle
\begin{abstract}
Retrieval-augmented generation (RAG) systems are increasingly deployed in sensitive domains such as healthcare and law, where they rely on private, domain-specific knowledge. This capability introduces significant security risks, including membership inference, data poisoning, and unintended content leakage. A straightforward mitigation is to enable all relevant defenses simultaneously, but doing so incurs a substantial utility cost. In our experiments, an always-on defense stack reduces contextual recall by more than 40\%, indicating that retrieval degradation is the primary failure mode. To mitigate this trade-off in RAG systems, we propose the Sentinel-Strategist architecture, a context-aware framework for risk analysis and defense selection. A Sentinel detects anomalous retrieval behavior, after which a Strategist selectively deploys only the defenses warranted by the query context. Evaluated across three benchmark datasets and five orchestration models, ADO is shown to eliminate MBA-style membership inference leakage while substantially recovering retrieval utility relative to a fully static defense stack, approaching undefended baseline levels. Under data poisoning, the strongest ADO variants reduce attack success to near zero while restoring contextual recall to more than 75\% of the undefended baseline, although robustness remains sensitive to model choice. Overall, these findings show that adaptive, query-aware defense can substantially reduce the security-utility trade-off in RAG systems.
\end{abstract}


\section{Introduction}
\label{sec:intro}
\begin{figure}[h]
    \centering
    \includegraphics[width=\linewidth]{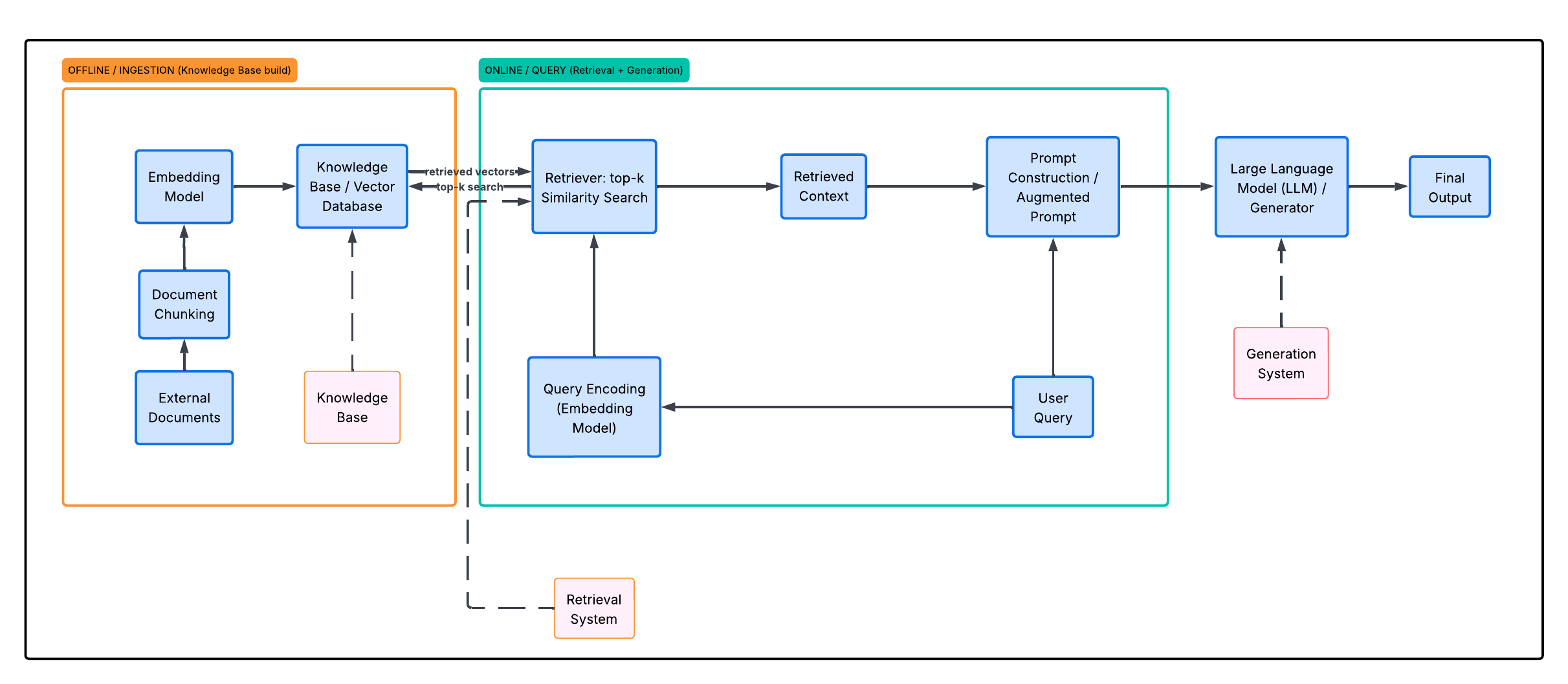}
    \caption{Architecture of the standard Retrieval-Augmented Generation (RAG) pipeline. The framework consists of three main phases: offline ingestion (document chunking and embedding to build the knowledge base), online retrieval (encoding the user query and performing a top-$k$ similarity search), and augmentation (prompt construction utilizing the retrieved context to condition the generator).}
    \Description{Block diagram of a standard RAG pipeline. Documents are ingested, chunked, and embedded into a vector knowledge base offline. At inference time, a user query is encoded, relevant documents are retrieved with top-$k$ similarity search, and the retrieved context is combined with the query before generation.}
    \label{ragpipeline}
\end{figure}

In the realm of Large Language Models (LLMs), significant advancements have been made across various tasks such as automated 
code generation, medical diagnostics, and document summarization \cite{ye2024cognitive,llm-hallucinations}. However, standalone 
LLMs suffer from two well-documented limitations: they tend to generate plausible but factually incorrect content, known as 
hallucinations, and they are limited by their training cutoff, unable to access information beyond it 
\cite{Ragsurvey,lewis2020retrieval,borgeaud2022improving}. Retrieval-Augmented Generation (RAG) directly addresses these issues 
by retrieving relevant passages from an external knowledge base before each generation step, thereby grounding responses in 
up-to-date, domain-specific evidence. As illustrated in Fig.~\ref{ragpipeline}, the standard RAG pipeline encodes a user query, 
retrieves the top-$k$ matching documents, and conditions the generator on the retrieved context. Due to these capabilities, RAG 
is widely adopted in high-stakes settings such as healthcare, legal analysis, and enterprise financial systems.

This integration of Large Language Models (LLMs) with a live, externally managed knowledge store in Retrieval-Augmented 
Generation (RAG) systems significantly expands the attack surface compared to standalone LLMs. RAG systems inherit 
vulnerabilities from standard LLMs while introducing new threats that target both the retrieval mechanism and the underlying 
knowledge base \cite{arzanipour2025ragsecurityprivacyformalizing}. These critical attack vectors pose significant concerns for 
deployed RAG systems:
\begin{itemize}
    \item \textbf{Membership Inference Attacks (MIAs)}, in which an adversary determines whether a specific sensitive document is 
present in the vector knowledge base by probing targeted queries \cite{Anderson_2025MIA,mba}.
    \item \textbf{Data Poisoning}, where malicious actors inject crafted adversarial documents into the knowledge store to make 
the retriever surface them in response to trigger queries and guide downstream generation towards attacker-controlled outputs 
\cite{zou2025poisonedrag}.
    \item \textbf{Content Leakage}, where the system reproduces verbatim or near-verbatim segments from private retrieved 
documents, exposing confidential information directly \cite{qi2024follow}.
\end{itemize}
The presence of these attack vectors poses a challenge to the core security properties of privacy, integrity, and confidentiality 
that RAG deployments must uphold \cite{arzanipour2025ragsecurityprivacyformalizing}.

While recent work has begun to formalize these threats, the predominant approach is still to study each attack vector in isolation. Recent surveys identify membership inference, content leakage, and data poisoning as core risks, yet fail to empirically quantify their joint impact on end-to-end RAG utility~\cite{arzanipour2025ragsecurityprivacyformalizing}. Attack and benchmark-focused work either targets a single class of adversary, such as membership inference against RAG \cite{ragleak,sma}, or concentrates on knowledge-base corruption and prompt-injection style poisoning without modeling privacy leakage \cite{zou2025poisonedrag,reliabilityrag,saferag}. To the best of our knowledge, we are not aware of prior empirical work that simultaneously (i) evaluates RAG under \emph{concurrent} multi-vector threats, specifically membership inference and data poisoning in our empirical study, while architecturally designing for content leakage, and (ii) measures the semantic utility cost of the defenses added to the system to tackle these attacks.

Existing defenses address individual stages of the pipeline: differentially private retrieval (DP-RAG) \cite{dp} perturbs query-document similarity scores \emph{inside the retriever} after encoding and before top-$k$ selection to suppress membership signals; TrustRAG-style clustering \cite{trustrag} filters semantic outliers from the retrieved set at the post-retrieval hook to neutralize poisoned documents; and attention-variance filters \cite{avfilters} inspect attention concentration over retrieved passages at the pre-generation hook to prune overly dominant context before decoding, thereby mitigating content leakage. A natural engineering response is to activate all three mechanisms simultaneously, which we term a \textbf{static full-defense stack}, so that every incoming query passes through all three layers regardless of the actual threat level. However, this indiscriminate combination imposes a cumulative overhead that disproportionately burdens the retrieval and conditioning stages: DP-RAG noise disrupts high-precision nearest-neighbour search, TrustRAG filtering aggressively prunes valid documents, and the attention-variance check conservatively prunes context that would otherwise ground accurate answers.

This leads to what we call the \textbf{security-utility paradox}: enabling the full defense stack reduces contextual recall by 41-46\% across benchmark datasets. Faithfulness, by contrast, remains stable, confirming that the LLM generator itself is unimpaired. The utility collapse originates largely in the retrieval phase: the generator is starved of sufficient context and can no longer meet its intended task requirements. Our experimental results in Section~\ref{sec:results} confirm this phenomenon empirically for the evaluated poisoning and membership-inference settings.

\begin{figure*}[t]
    \centering
    \includegraphics[width=\textwidth]{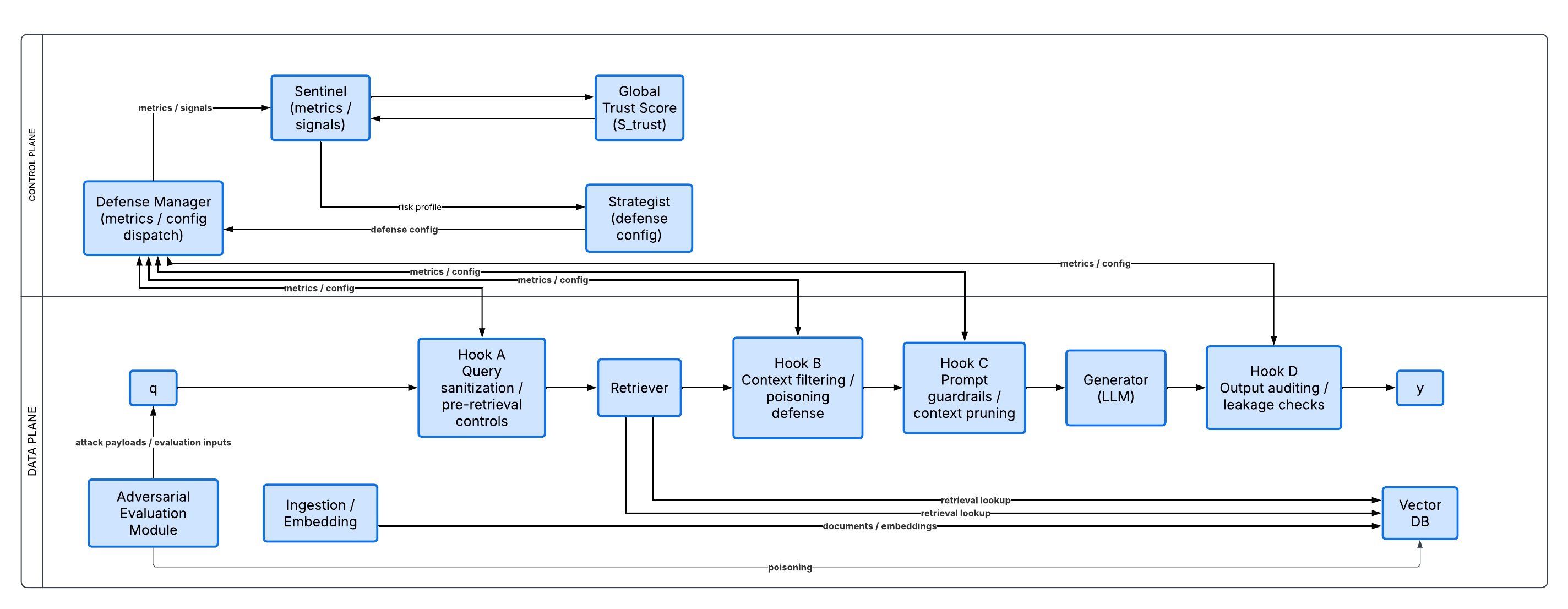}
    \caption{The Sentinel-Strategist Architecture for Adaptive Defense Orchestration (ADO). The system is divided into a Control Plane and a Data Plane. The Control Plane acts as a centralized Policy Decision Point: the Sentinel ingests per-query signals and pipeline metrics to produce a structured risk profile, which is passed to the Strategist to generate a defense configuration via the Defense Manager. The Data Plane applies those configurations across four enforcement hooks in the RAG pipeline, Hook A (query sanitization and pre-retrieval controls), Hook B (context filtering and poisoning defense), Hook C (prompt guardrails and context pruning), and Hook D (output auditing and leakage checks). The Adversarial Evaluation Module and Ingestion pipeline interface with the query path and vector database to support red-teaming, poisoning injection, and corpus updates.}
    \Description{Two-plane diagram of the Sentinel-Strategist ADO architecture. The control plane contains the Sentinel, the global trust score, the Strategist, and the Defense Manager, which exchange metrics and defense configurations. The data plane routes query q through Hooks~A-D, the retriever, the generator, and output y, with connections to the adversarial evaluation module, ingestion and embedding, and the vector database.}
    \label{ProposedPipeline}
\end{figure*}

To resolve this paradox, we propose \textbf{Adaptive Defense Orchestration (ADO)}, a modular framework that separates risk assessment from defense enforcement. Rather than applying all defenses to every query, ADO activates only the mechanisms warranted by the current threat level, preserving retrieval quality for benign workloads while strengthening protection against the evaluated threats when adversarial signals are detected. We realize this policy through the \textbf{Sentinel-Strategist architecture}, a two-stage orchestrator shown in Fig.~\ref{ProposedPipeline}. The split is intentional rather than cosmetic: threat assessment and defense selection have different inputs, objectives, and failure modes. The Sentinel continuously monitors lightweight pipeline signals, including lexical query overlap and vector-space dispersion among retrieved documents, and compresses them into a structured per-query risk profile. The Strategist then maps that profile to targeted defense configurations across four enforcement hooks in the RAG pipeline, dynamically enabling or tightening individual defenses as warranted. Decoupling evidence fusion from action selection makes the control policy easier to audit and recalibrate, allows the defense registry to evolve without rewriting the detector, and keeps the control prompts small enough to run on compact controller models. This improves practical deployment viability, preserves retrieval performance on benign traffic, and provides an adaptive security posture.

Conceptually, this design aligns with the Zero Trust Architecture (ZTA) model formalized by NIST~SP~800-207~\cite{nist_800_207}, in which a centralized Policy Decision Point (PDP) continuously evaluates contextual risk and instructs distributed Policy Enforcement Points (PEPs) that gate access to protected resources.

\noindent Our contributions are as follows:
\begin{itemize}
    \item We provide an empirical study of RAG systems under two evaluated attack classes, membership inference and data poisoning, and quantify the semantic utility cost of statically combining the corresponding defenses.
    \item We demonstrate that stacking the available defense modules as an always-on static policy causes a 41-46\% collapse in contextual recall, even when the underlying generator remains fully intact, identifying retrieval as the primary bottleneck.
    \item We introduce ADO and the Sentinel-Strategist architecture, a unified multi-vector orchestration framework that coordinates defenses for membership inference, data poisoning, and content leakage through separate enforcement hooks. The Sentinel performs lightweight risk estimation, while the Strategist maps that risk profile to hook-level actions; this separation makes the policy more auditable, easier to extend, and practical to run with compact controller models. In this paper, the empirical evaluation centers on the first two attack classes, while content leakage is incorporated architecturally through the AV-filter path.
    \item We provide code, attack configurations, Sentinel and Strategist prompt templates, and evaluation scripts to support independent reproducibility; submission-time artifact access is described in the Open Science appendix.
\end{itemize}

\section{Related Work}
\subsection{Retrieval-Augmented Generation (RAG)}

Retrieval-Augmented Generation (RAG) is designed to increase the capabilities of Large Language Models (LLMs) by combining them with external knowledge sources \cite{lewis2020retrieval}. This approach effectively handles the issues of hallucinations and knowledge cutoff limitations found in standalone parametric models \cite{shuster-etal-2021-retrieval-augmentation,gao2023retrieval}. In contrast to traditional LLMs that depend solely on fixed internal parameters, RAG systems actively fetch semantically relevant context from external databases. As illustrated in Fig.~\ref{ragpipeline}, the pipeline consists of three 
phases: ingestion (document chunking and embedding), retrieval (top-$k$ similarity search), and augmentation (prompt construction for the generator)~\cite{lewis2020retrieval}. Recent developments have progressed to include advanced techniques such as token-level retrieval methods \cite{khandelwal2019generalization}, adaptive data chunking strategies \cite{ram-etal-2023-context}, and graph-structured knowledge models that represent complex relational dependencies among entities \cite{edge2025localglobalgraphrag,wang-etal-2025-knowledge-graph}.

\subsection{Privacy \& Security Risks in RAG Systems}
Integrating external knowledge bases significantly improves LLMs, but it also changes the threat landscape by creating additional attack surfaces. Therefore, we categorize these risks into three primary vectors: membership inference, data poisoning, and content leakage.

\subsubsection{Membership Inference Attacks (MIA)}
Membership Inference Attacks (MIA) \cite{mia1,mia2} pose a privacy risk by attempting to determine whether a specific data entry was included in the training dataset or knowledge base of a machine learning model. For conventional language models, MIAs have been thoroughly examined, with notable techniques such as the Loss Attack, which infers membership based on the model's loss \cite{yeom2018privacy}, the Zlib Entropy Attack, which adjusts the loss by considering compression size \cite{carlini2021extracting}, and the Min-\(k\)\% Probability Attack, which uses the least probable tokens in a sample to assess membership \cite{shi2023detecting}. RAG-MIA \cite{Anderson_2025MIA} determines membership by directly querying the system to identify whether a specific document is included in the retrieved context, based on the model's response. Extending this to a more targeted setting, the Mask-Based Attack (MBA) \cite{mba} strategically masks tokens within a candidate document and queries the RAG system to assess whether it reconstructs the masked content; successful reconstruction indicates that the document is present in the retrieval knowledge base, making MBA particularly effective against systems that retrieve and expose verbatim document segments.

\subsubsection{Data Poisoning}
Data poisoning attacks targeting Retrieval-Augmented Generation (RAG) systems \cite{zou2025poisonedrag,zhang2025practical} manipulate model outputs by inserting adversarial documents into the retrieval knowledge base while leaving the model parameters unchanged. An attacker adds specially crafted documents (\(D_{\text{poi}}\)) to the original corpus (\(D\)), thereby creating a contaminated database \(D^{\star} = D \cup D_{\text{poi}}\). This ensures that the retriever yields these documents in response to particular trigger queries. Data poisoning attacks typically fall into two categories: harmful generation and content promotion. In harmful generation, the attacker aims to produce misleading, biased, or dangerous outputs. In content promotion, the goal is to steer the system toward specific target phrases, entities, or brand mentions that may not be relevant to the user's intent. A poisoning attack is considered successful if the retriever yields at least one injected document in response to a trigger query \(q^{*}\):
\begin{equation}
R(\mathcal{E}(q^{*}), \mathcal{E}(D^{\star}), k) \cap D_{\text{poi}} \neq \emptyset.
\end{equation}

\subsubsection{Content Leakage and Unintended Disclosure}
In Retrieval-Augmented Generation (RAG) systems, content leakage refers to the unauthorized reconstruction of sensitive information from the retrieval knowledge base. This risk is particularly critical in domains such as healthcare, where the generator (\(G\)) may produce segments that are similar to retrieved documents (\(d\)), thereby exposing patient information. Adversaries can induce such leakage using a combined query approach \(q = q_i + q_c\). An anchor query (\(q_i\)) steers the retriever (\(R\)) toward a specific group of sensitive documents, while a command prompt (\(q_c\)) instructs the generator to replicate the retrieved content with minimal modifications. Leakage occurs when the similarity between the generated output (\(y\)) and a sensitive document (\(d_i\)) exceeds a predefined threshold \(\tau\). Recent studies \cite{arzanipour2025ragsecurityprivacyformalizing} systematically investigate unauthorized content leakage in RAG systems and its privacy implications.

\subsection{Defenses}

Given the growing threat to RAG systems, specialized defense mechanisms are required. All these strategies target specific stages of the RAG pipeline, like retrieval, augmentation, or generation, to reduce distinct attack vectors.

\subsubsection{Defenses Against Membership Inference}

In Membership Inference Attacks (MIAs), an attacker tries to find if a particular record exists within a system's database. Recent studies apply Differential Privacy (DP) \cite{dp_math} to the retrieval stage of Retrieval-Augmented Generation (RAG) systems \cite{dp}. A common instantiation perturbs query-document similarity scores after query and document encoding but before final top-$k$ ranking, thereby reducing the influence of any single document on the retrieved set and limiting the adversary's capacity to identify its presence.

A randomized mechanism $\mathcal{M}$ is said to fulfill $(\varepsilon,\delta)$-differential privacy if, for every pair of neighboring datasets $D$ and $D^{-}$ that vary by exactly one entry, as well as for any measurable output set $\mathcal{O}$, the following condition is satisfied:
\begin{equation}
\Pr[\mathcal{M}(D) \in \mathcal{O}] \leq e^{\varepsilon} \Pr[\mathcal{M}(D^{-}) \in \mathcal{O}] + \delta .
\end{equation}

\subsubsection{Defenses Against Data Poisoning}

The main focus of Defenses against data poisoning is post-retrieval filtering and robust aggregation before it reaches the generation phase. The goal is to recognize and eliminate malicious documents, such as those that insert misleading statements or specific keywords, which the retriever pulls in along with valid evidence. TrustRAG \cite{trustrag} implements a multi-stage validation process to enhance the reliability of the system, based on the idea that poisoned documents frequently appear as semantic anomalies. ReliabilityRAG \cite{reliabilityrag} enhances by prioritizing resilience for RAG systems. In contrast to heuristic filtering, ReliabilityRAG offers statistical assurances that the system's outputs will remain consistent, even when there are a limited number of harmful documents among the top-k results.

\subsubsection{Defenses Against Content Leakage \& Unintended Disclosure}

Measures to prevent content leakage concentrate on the safety barrier during the generation process. Their main goal is to stop the Large Language Model (LLM) from reproducing exact or near-exact sections of gathered documents that may contain confidential or sensitive information, such as personally identifiable information in medical records. ControlNet~\cite{controlnet} is presented as a firewall for retrieval-augmented generation (RAG) systems. Rather than depending on keyword filtering, ControlNet observes the model's internal activation patterns and hidden states throughout the generation process. In the Attention-Variance Filter \cite{avfilters}, the generator's attention over the retrieved context is monitored for anomalously concentrated weights. When a model is compelled to disclose certain content, often through malicious prompting or poisoning, it assigns a disproportionately high attention weight to specific tokens within the passage to override its internal biases.

\section{System Model and Threat Landscape}
In this section, we outline the retrieval-augmented generation (RAG) system model, the threat landscape, the security-utility paradox, and the adversarial capabilities. We first define the underlying RAG pipeline, then describe the adversary's interaction model and objectives, and finally state the security-utility problem that motivates our defense architecture. 

\subsection{Retrieval-Augmented Generation (RAG) System Model}
As shown in Fig.~\ref{ragpipeline}, the RAG system comprises three main components: a knowledge base, a retrieval system, and a generation system. A large language model (LLM) powers the generation component, while the knowledge base and retriever work together to select and supply contextual information. This information conditions the model's output on relevant and factual data.

\subsubsection{Offline Ingestion Phase}
Prior to inference, the knowledge corpus must be prepared to enable efficient retrieval. The underlying document collection is partitioned into a set of discrete passages $D = \{d_1, \ldots, d_N\}$. The encoder $\mathcal{E}$ computes dense vector representations for all passages, yielding the encoded corpus $\mathcal{E}(D)$, which is stored in a vector index.

\subsubsection{RAG Inference Pipeline}
Throughout the manuscript, we use the retriever signature
\[
D_q = R\!\left(\mathcal{E}(q),\,\mathcal{E}(D),\,k\right),
\]
where $R(\cdot)$ denotes top-$k$ retrieval over the encoded corpus. At inference time, the encoder embeds query $q$, the retriever selects top-$k$ documents $D_q$, and the generator $G$ produces response $y$ conditioned on the augmented prompt:
\[
y = G\!\left(\mathrm{Augment}\!\left(q,\; 
    R\!\left(\mathcal{E}(q),\,\mathcal{E}(D),\,k\right),\; s\right)\right).
\]
where $s$ denotes system-level instructions, and $\mathrm{Augment}(\cdot)$ is a deterministic concatenation operation.
\subsection{Adversarial Model}
~\label{adversarial_model}

\subsubsection{Adversary Knowledge}

We consider a black-box adversary \(\mathcal{A}\) that observes only query-response pairs. Formally, given the system \(\mathcal{S}\), the adversary can choose a sequence of queries
\[
q_1, q_2, \ldots, q_T
\]
and observe the corresponding outputs
\[
\begin{aligned}
y_t &= G\!\left(\mathrm{Augment}\!\left(q_t,\, R\!\left(\mathcal{E}(q_t), \mathcal{E}(D), k\right),\, s\right)\right), \\
&\qquad t = 1, \ldots, T.
\end{aligned}
\]

The adversary has \emph{no} direct access to the generator $G$'s model parameters $\theta$, to its internal activations or gradients, or to the raw documents in the corpus $D$. We also assume that the adversary does not control the embedding function $\mathcal{E}$ or the retrieval index implementation; the defender operates these.

\subsubsection{Adversary Capabilities}

Within this black-box setting, the adversary has two main capabilities.

\paragraph{Query Access}
The adversary can issue an unbounded (or practically high) number of queries to the system, possibly at machine speed. We denote by \(\mathcal{Q}\) the space of admissible queries, and write \(Q = \{q_t\}_{t=1}^{T} \subset \mathcal{Q}\) for the set of queries submitted during an attack episode. This capability underlies both probing for membership information and crafting prompts that induce leakage.

\paragraph{Data Injection.}
In addition, we allow the adversary to tamper with the knowledge base by injecting a set of crafted documents \(D_{\text{poi}}\) into the ingestion pipeline. The resulting corpus becomes
\[
D^{\star} = D \cup D_{\text{poi}},
\]
and the retriever subsequently operates over the encoded corpus \(\mathcal{E}(D^{\star})\). This capability models a realistic compromise of upstream data sources or ingestion jobs in RAG deployments.

We do not consider stronger capabilities such as direct modification of model weights, exfiltration of the entire corpus \(D\), or control over the embedding and retrieval infrastructure. These are out of scope for this work.

\subsubsection{Threat Objectives}

The adversary's behavior has three distinct objectives, corresponding to standard security properties in information forensics: privacy, integrity, and confidentiality. We formalize each of these objectives below.

\paragraph{Membership Inference (Privacy Violation).}

The adversary seeks to determine whether the knowledge base \(D\) contains a specific sensitive document \(d_{\text{target}}\). To make the non-member case explicit, we define the adjacent corpus
\[
D^{-} = D \setminus \{d_{\text{target}}\},
\]
and let \(\mathcal{A}\) issue a set of membership probes \(Q_{\text{mia}}\). In our setting, this objective is instantiated with the Mask-Based Attack (MBA): each probe masks selected tokens from \(d_{\text{target}}\) and queries the RAG system to reconstruct them. Let \(r(q; D) \in \{0,1\}\) denote whether probe \(q \in Q_{\text{mia}}\) exactly reconstructs the masked target span when the system operates over corpus \(D\). A membership signal exists when exact reconstruction is more likely under the member corpus than under the adjacent non-member corpus, i.e.,
\[
\Pr\!\left[r(q; D)=1\right] > \Pr\!\left[r(q; D^{-})=1\right].
\]
This formulation keeps the underlying objective membership-based while matching the reconstruction-style signal produced by MBA. Because our empirical study does not threshold that signal into a separate binary detector, we operationalize MIA empirically using MBA exact mask-fill leakage rate:
\[
L_{\text{mia}} = \frac{1}{|Q_{\text{mia}}|} \sum_{q \in Q_{\text{mia}}} r(q; D).
\]
That is, the reported MIA leakage rate is the percentage of evaluation probes for which the masked target content is reconstructed exactly as the reference text; in Section~\ref{sec:experiments}, we report it either in aggregate or separately for member and non-member probe sets.

\paragraph{Data Poisoning (Integrity Violation).}

In the poisoning setting, the adversary injects a set of malicious documents \(D_{\text{poi}}\) into the corpus and aims to steer the system towards a targeted adversarial output \(y_{\text{adv}}\) for a specific trigger query \(q_{\text{trig}}\). After poisoning, the effective corpus is \(D^{\star} = D \cup D_{\text{poi}}\), and the retriever operates as \(R(\mathcal{E}(q), \mathcal{E}(D^{\star}), k)\).

The attacker's objective is to maximize
\[
\Pr\!\left(
y = y_{\text{adv}}
\;\middle|\;
q_{\text{trig}},
R\big(\mathcal{E}(q_{\text{trig}}), \mathcal{E}(D^{\star}), k\big)
\right).
\]
\paragraph{Content Leakage (Confidentiality Violation).}

For content leakage, the adversary constructs a query \(q_{\text{leak}}\) intended to coerce the generator into revealing verbatim or near-verbatim content from private documents. Let \(D_{\text{priv}} \subseteq D\) represent a subset of sensitive documents that contains personal information, with \(d_{\text{priv}} \in D_{\text{priv}}\) signifying a particular target document.

The attack succeeds if the similarity between the generated output \(y\) and \(d_{\text{priv}}\) exceeds a predefined threshold \(\tau\):
\[
\text{sim}\big(y, d_{\text{priv}}\big) > \tau.
\]
While we formalize content leakage here to provide a complete view of the RAG threat landscape, our empirical evaluation in Section~\ref{sec:experiments} restricts its attack benchmarks to membership inference and data poisoning, including their joint evaluation on TriviaQA.

\paragraph{Concurrent Attacks.}

We allow the adversary to pursue these objectives concurrently over time. In other words, a single adversarial client may interleave membership probes, poisoning triggers, and leakage prompts within the same query stream. Consequently, any practical defense must operate under a unified orchestration framework that balances robustness against all three threat vectors with the preservation of system utility, rather than tuning to a single attack in isolation.

\subsection{Security-Utility Problem Statement}
~\label{ps}

\subsubsection{Defense Policies}

We model a defense configuration as a policy \(\pi\) that, for each incoming query \(q\) and system state \(s\), selects a setting for the available defenses. Concretely, \(\pi\) can enable or disable mechanisms such as differentially private retrieval (DP-RAG), TrustRAG-style consistency filtering, and attention-variance-based leakage filters, and adjust their hyperparameters (e.g., privacy budget \(\epsilon\) or similarity thresholds).

Let \(\Pi\) denote the space of admissible policies. For a given policy \(\pi \in \Pi\), query \(q\), and underlying corpus \(D\), the defended system induces a distribution over outputs
\[
y \sim \mathcal{S}^{\pi}(q, D),
\]

\subsubsection{Utility and Risk Measures}

To quantify semantic utility, we employ an LLM-as-a-judge framework\cite{llmjudge} to compute a vector of evaluation metrics. The judge evaluates the system using combinations of four core parameters: the user query $q$, the retrieved context $D_q$, the expected reference output $y^*$, and the actual generated output $y$. Let $J_{\text{LLM}}$ denote the evaluation function. Specifically, we compute four distinct metrics, each relying on a specific subset of these parameters:
\begin{itemize}
    \item \textbf{Contextual Recall:} Evaluates if the retrieved context covers the expected answer, requiring $(y^*, D_q)$ \footnote{https://deepeval.com/docs/metrics-contextual-recall\#how-is-it-calculated}.
    \item \textbf{Contextual Relevancy:} Measures the relevance of the retrieved context to the query, requiring $(q, D_q)$ \footnote{https://deepeval.com/docs/metrics-contextual-relevancy\#how-is-it-calculated}.
    \item \textbf{Answer Relevancy:} Assesses how directly the generated answer addresses the query, requiring $(q, y)$ \footnote{https://deepeval.com/docs/metrics-answer-relevancy\#how-is-it-calculated}.
    \item \textbf{Faithfulness:} Verifies that the generated answer is strictly grounded in the retrieved context, requiring $(D_q, y)$
    \footnote{https://deepeval.com/docs/metrics-faithfulness\#how-is-it-calculated}.
\end{itemize}
Consequently, the overall utility vector can be formalized as:
\[
\mathcal{U}(q, y^*; \pi) = J_{\text{LLM}}(q, D_q, y^*, y) \in \mathbb{R}^4,
\]
where $\pi$ denotes the system policy that induces $D_q$ and $y$.

In parallel, we define a risk vector representing vulnerabilities to specific attacks:
\[
\mathcal{R}(q; \pi) = \big(R_{\text{mia}}(q; \pi),\, R_{\text{poi}}(q; \pi),\, R_{\text{leak}}(q; \pi)\big),
\]
where the components correspond to the risk of membership inference, data poisoning, and content leakage, respectively.

\subsubsection{Security-Utility Optimization}

At a high level, the defender seeks a policy \(\pi\) that maximizes expected utility for benign workloads while keeping expected risk under specified tolerances. This can be expressed as the constrained optimization problem
\[
\max_{\pi \in \Pi} \; \mathbb{E}_{q \sim \mathcal{Q}_{\text{benign}}}
\big[ u(\mathcal{U}(q; \pi)) \big]
\quad \text{s.t.} \quad
\bar{\mathcal{R}}(\pi) \preceq \boldsymbol{\tau},
\]
where \(u(\cdot)\) is a scalar aggregation of the utility vector (e.g., a weighted sum of contextual recall and answer relevancy), \(\boldsymbol{\tau}\) is a vector of acceptable risk thresholds for membership inference, poisoning, and leakage, and \(\preceq\) denotes element-wise inequality.

In practice, we evaluate a small set of representative policies rather than solving this optimization exactly. Of particular interest are:


We evaluate three representative policies: an undefended 
$\pi_{\text{base}}$, a fully-armed static stack 
$\pi_{\text{static}}$, and the proposed adaptive 
$\pi_{\text{ado}}$, described in Section~\ref{methods}.

\subsubsection{The Security-Utility Paradox}

We define the security-utility paradox in RAG systems as the empirical phenomenon in which defensive policies such as \(\pi_{\text{static}}\) satisfy the risk constraints \({\mathcal{R}}(\pi_{\text{static}}) \preceq \boldsymbol{\tau}\), but drive the utility term \({\mathcal{U}}(\pi_{\text{static}})\) to low values. In our evaluation, this results in a significant reduction in contextual recall on the order of \(41-46\%\) across datasets. However, faithfulness remains unchanged, indicating that the underlying generator retains its reasoning ability but no longer receives adequate context.

Our work addresses this problem along two complementary axes. First, we introduce a utility-aware evaluation framework that jointly quantifies semantic utility and adversarial risk under different defense policies, making the security-utility trade-off explicit. Second, we propose an Adaptive Defense Orchestration (ADO) policy \(\pi_{\text{ado}}\), realized via the Sentinel-Strategist architecture, which dynamically configures defenses per query in an attempt to preserve utility while maintaining strong protection against the threats in Section~\ref{adversarial_model}. Section~\ref{methods} details the design of ADO and the underlying orchestration mechanisms, while Section~\ref{sec:experiments} describes the experimental protocol used to instantiate and evaluate the proposed framework.

\section{Proposed Methodology: Adaptive Defense Orchestration}
~\label{methods}
In this section, we present the Adaptive Defense Orchestration (ADO) framework, which operationalizes the security-utility optimization objective defined in Section~\ref{ps}. Rather than treating individual attacks or defenses in isolation, ADO provides both (i) a hook-based Data Plane that exposes measurable enforcement points for systematically quantifying the trade-off between security and semantic utility, and (ii) a stateful Control Plane, realized through the Sentinel-Strategist orchestrator, that dynamically configures defenses per query in response to observed risk.

\subsection{Architectural Overview}

ADO wraps the base RAG system $\mathcal{S} = (\mathcal{E}, R, G, D)$, with a \textbf{Control Plane} and a \textbf{Data Plane}, following a strict separation of policy decision from policy enforcement, a design that directly instantiates the Zero Trust Architecture (ZTA) model of NIST SP~800-207~\cite{nist_800_207}, in which a centralized Policy Decision Point (PDP) continuously evaluates contextual risk and instructs distributed Policy Enforcement Points (PEPs) that gate access to protected resources.

The \textbf{Control Plane} (PDP) comprises the Sentinel risk assessment stage, the Strategist defense configuration stage, and the Persistence Layer maintaining the per-user Global Trust Score $S_{\text{trust}}$. It is responsible exclusively for deciding \textit{which} defenses to activate and \textit{how} to parametrize them per query, without touching query or document content directly. Section~\ref{sso} details its internal operation.

The \textbf{Data Plane} (PEPs) comprises the Core RAG Engine and the four interception hooks pre-retrieval, post-retrieval, pre-generation, and post-generation, that wrap it. It handles all live query traffic and executes the defense configurations issued by the Control Plane. Section~\ref{dframework} describes these hook-level mechanisms and the Adversarial Evaluation Module that enables controlled red-teaming against this layer.

\subsection{Data Plane}
\label{dframework}
To enable both systematic evaluation and dynamic control of defenses, we wrap the base RAG system \(\mathcal{S} = (\mathcal{E}, R, G, D)\) with four configurable interception hooks. The Core RAG Engine forms the execution backbone of the framework. It implements the end-to-end RAG pipeline, including data ingestion, document chunking, embedding into the vector database, query encoding, retrieval, prompt augmentation, and response generation. 

\subsubsection{Interception Hooks}

Around the Core RAG Engine, the Data Plane uses lightweight interceptor-style middleware that provides a unified interface for enabling, disabling, or reconfiguring security mechanisms at query runtime. Through this wrapper, ADO exposes four interception hooks aligned with key stages of the RAG pipeline:

\begin{itemize}
    \item \textbf{Pre-Retrieval Hook.} Intercepts the raw query \(q\) prior to encoding. This hook supports query sanitization (e.g., removal of personally identifiable information), benign transformations (e.g., query expansion), and the injection of prompt-level guardrails before the query enters the embedding function \(\mathcal{E}\).
    \item \textbf{Post-Retrieval Hook.} Intercepts the retrieved document set \(D_q = R(\mathcal{E}(q), \mathcal{E}(D), k)\). At this stage, the framework can apply defenses such as TrustRAG-style clustering and consistency checks to identify and remove semantic outliers, or other post-retrieval filters designed to neutralize poisoning attempts. Algorithm~\ref{alg:trustrag} summarizes the TrustRAG filtering procedure used at this hook.
    \item \textbf{Pre-Generation Hook.} Wraps the augmented prompt \(q'\) that will be passed to the generator \(G\). This hook is used to apply system-level safety prompts, jailbreak- and prompt-injection defenses, and attention-variance-based context pruning that removes disproportionately dominant passages before decoding. Algorithm~\ref{alg:avfilter} provides the corresponding attention-variance pruning routine.
    \item \textbf{Post-Generation Hook.} Audits the final output \(y\) against both the query \(q\) and the retrieved context \(D_q\). Here, we attach output-side mechanisms such as hallucination detectors or policy-based response filters, which can mask or regenerate tokens when the generated answer deviates from grounded evidence or violates disclosure constraints.
\end{itemize}

For reproducibility, we include pseudocode for the two representative hook-level mechanisms discussed above.

\begin{algorithm}[!t]
\caption{TrustRAG: Clustering-Based Poisoning Filter}
\label{alg:trustrag}
\begin{algorithmic}[1]
\Require Retrieved documents $D_q$ with embeddings $E$, Similarity threshold $\sigma$, ROUGE threshold $\rho$
\Ensure Filtered document set $D_f \subseteq D_q$

\State Normalise embeddings in $E$
\State $(C_0, C_1) \leftarrow \textsc{KMeans}(E,\ k{=}2)$
\State Compute intra-cluster cosine means $\overline{s}_0,\ \overline{s}_1$ for $C_0, C_1$

\If{$\min(\overline{s}_0, \overline{s}_1) \geq \sigma$}
    \State $D_f \leftarrow D_q$ \Comment{Both clusters are coherent; retain all}
\Else
    \If{$\overline{s}_0 \geq \overline{s}_1$}
        \State $D_f \leftarrow C_0$
    \Else
        \State $D_f \leftarrow C_1$
    \EndIf
    \State \Comment{Retain cluster with higher coherence}
\EndIf

\If{$|D_f| > 1$}
    \ForAll{$d \in D_f$}
        \State $c(d) \leftarrow \frac{1}{|D_f|-1} \sum\limits_{d' \in D_f,\ d' \neq d} \cos(d,d')$
    \EndFor

    \ForAll{unordered pairs $(d_i, d_j) \subseteq D_f$}
        \If{$\textsc{ROUGE-L}(d_i, d_j) > \rho$}
            \State Remove $\arg\min\limits_{d \in \{d_i,d_j\}} c(d)$ from $D_f$
            \Comment{Prune near-duplicate with lower coherence}
        \EndIf
    \EndFor
\EndIf

\State \Return $D_f$
\end{algorithmic}
\end{algorithm}

\begin{algorithm}[!t]
\caption{Attention-Variance Filter: Context Pruning}
\label{alg:avfilter}
\begin{algorithmic}[1]
\Require Context passages $C = \{c_1, \ldots, c_n\}$, question $q$, maximum removals $R$, variance threshold $\vartheta$
\Ensure Pruned context $C' \subseteq C$

\State $C' \leftarrow C$
\For{$r = 1$ \textbf{to} $R$}
    \State $p \leftarrow \textsc{NormalizedAttention}(q, C')$
    \If{$\mathrm{Var}(p) \leq \vartheta$}
        \State \textbf{break}
        \Comment{Attention is evenly distributed; no anomalous passage.}
    \EndIf
    \State $j^* \leftarrow \arg\max_{j}\ p_j$
    \State $C' \leftarrow C' \setminus \{c_{j^*}\}$
    \Comment{Remove passage with disproportionately high attention}
\EndFor

\State \Return $C'$
\end{algorithmic}
\end{algorithm}

We implement each concrete defense as a pluggable module. This includes differentially private retrieval (DP-RAG), TrustRAG-style post-retrieval filtering, and attention-variance-based pre-generation context pruning. These modules can be bound to one or more hooks with configurable parameters (e.g., privacy budget \(\epsilon\), clustering thresholds, similarity cutoffs). This design allows the Sentinel-Strategist orchestrator to reconfigure the security posture by adjusting hook-level parameters, without requiring changes to the Core RAG Engine.

\subsubsection{Adversarial Evaluation Module}

To support rigorous, repeatable evaluation of security-utility trade-offs under different defense policies, we introduce an Adversarial Evaluation Module that acts as an automated red-teaming environment. This module interacts with the framework in an offline or batch setting and comprises three main components:

\paragraph{Poison Injector.} The Poison Injector targets the ingestion layer by injecting a controlled set of adversarial documents \(D_{\text{poi}}\) into the knowledge base. These documents are designed with particular trigger phrases or semantic similarities so that, according to the retrieval function \(R\), they show up among the top-\(k\) results for specific trigger queries \(q_{\text{trig}}\). By varying the number and content of injected documents, we can systematically study the susceptibility of different defense policies to data poisoning and its impact on retrieval accuracy and downstream outputs.

\paragraph{Adversarial Payload Generator and Probe Construction.} The Adversarial Payload Generator constructs the adversarial query sets used to probe privacy and confidentiality risks. It programmatically generates membership inference probes, leakage-inducing prompts, and other attack payloads, together with the metadata needed for evaluation such as identifiers of target documents, mask positions, or poisoning keywords. This component ensures that the same adversarial workloads can be replayed across policies \(\pi_{\text{base}}, \pi_{\text{static}}, \pi_{\text{ado}}\) for fair comparison.

\paragraph{Automated Evaluator and Metric Computation.} The Automated Evaluator closes the loop by consuming paired datasets of adversarial (and benign) payloads and corresponding system outputs. For each policy \(\pi\), it computes the security and utility metrics defined in Section~\ref{sec:experiments}, including attack success rate for poisoning, MBA exact mask-fill leakage rates on member and non-member probe sets, leakage similarity statistics, and semantic utility scores such as contextual recall, contextual relevancy, answer relevancy, and faithfulness. These measurements instantiate the abstract utility \({\mathcal{U}}(\pi)\) and risk \({\mathcal{R}}(\pi)\) objectives from Section~\ref{ps}, enabling direct empirical comparison between static full-defense configurations and the proposed adaptive orchestration.

\subsection{Control Plane: Sentinel-Strategist Orchestration}
\label{sso}

The Control Plane is the Sentinel-Strategist orchestration logic, 
serving as a centralized Policy Decision Point (PDP). It observes 
pipeline telemetry from the Data Plane and issues per-query defense 
configurations back to the enforcement hooks without executing the core 
retrieval or generation path. In our implementation, the Control Plane does inspect query-side information and derived retrieval metrics for policy decisions, but it does not consume raw document text from the knowledge base.

The orchestration logic is implemented in three components: a persistence layer that maintains a per-user Global Trust Score, the Sentinel risk assessment stage, and the Strategist defense configuration stage.

In our method, both the Sentinel and the Strategist are instantiated as large language model (LLM) agents, but with deliberately different roles. The Sentinel takes as input a structured description of the current query, user state, and pipeline metrics, and outputs a risk profile. The Strategist consumes only that profile together with the current trust state and produces a defense plan for the four enforcement hooks. We separate these stages because risk estimation and defense selection require different reasoning: the Sentinel aggregates weak signals and preserves calibration, whereas the Strategist operates over a constrained action space that should remain stable even as the defense registry evolves. This modularity also leaves open future deployment variants in which the Strategist could be partially static or rule-based once the mapping from risk profiles to defense actions has been calibrated. A monolithic controller would entangle detection with actuation, enlarge the prompt, and make failures harder to diagnose. The split instead yields short, structured prompts that can be served by smaller control models, reducing per-query latency and cost. Operationally, this yields a two-pass LLM pipeline per query: an early pass before retrieval to configure pre-retrieval defenses, and a second pass after retrieval to refine the risk assessment using derived retrieval metrics and to configure post-retrieval and post-generation defenses.

\subsubsection{Persistence Layer and Global Trust Score}

In order to address slow-burn attacks where attackers spread their probing or poisoning efforts across numerous queries, the system maintains a Global Trust Score \(S_{\text{trust}} \in [0,1]\) for each user. Conceptually, a user with a score of \(S_{\text{trust}} = 1.0\) has a record of safe, valuable interactions. A score of \(S_{\text{trust}} = 0.0\) signifies a history of suspicious behavior, such as repeated jailbreak attempts or frequent activation of leakage filters. After each interaction, based on observed signals, the Global Trust Score is updated. The observed signals include whether defenses were triggered, whether retrieved context had to be pruned before generation, and whether outputs were masked or regenerated by output-side audits. \(S_{\text{trust}}\) serves as a prior in the orchestration logic. Lower trust values reduce the activation threshold for strict defenses, which makes them easier to trigger and thereby increases scrutiny. Higher trust values raise the threshold, resulting in a lighter default security posture.

Formally, after each interaction, the Sentinel emits a score delta $\delta \in [-0.1, 0.1]$ at both the pre-retrieval and post-retrieval passes, based on whether the query exhibited adversarial patterns, whether defenses were triggered, whether context pruning was required, and whether outputs were masked or regenerated by downstream audits. The Global Trust Score is then updated as:

\begin{equation}
    S_{\text{trust}}^{(t+1)} = \text{clip}\!\left(S_{\text{trust}}^{(t)} 
    + \delta_{\text{pre}} + \delta_{\text{post}},\; 0,\; 1\right)
    \label{eq:trust_update}
\end{equation}

\noindent where $\delta_{\text{pre}}$ and $\delta_{\text{post}}$ are the deltas emitted by the Sentinel at the pre- and post-retrieval passes respectively, and $\text{clip}(\cdot, 0, 1)$ ensures $S_{\text{trust}}$ remains a valid probability. A benign interaction (no defense activations, no context pruning, no masked output) yields a positive $\delta$, gradually raising trust. A suspicious interaction yields a negative $\delta$, lowering the activation threshold for subsequent defenses. This design addresses slow-burn attacks in which adversaries spread probing attempts across many queries to evade per-query detectors.

\subsubsection{Stage 1: Sentinel Risk Assessment}

To estimate the current threat level, the Sentinel acts as an intelligence fusion center that aggregates evidence from multiple sources. For each incoming query \(q\), it takes four categories of input: the Global Trust Score \(S_{\text{trust}}\) for the requesting user, the raw query text \(q\), a configurable window of recent session history (past queries and defense activations), and a set of computationally inexpensive metrics computed over the query and retrieved results. 

Before activating the generator \(G\), the system calculates a set of heuristic indicators that anchor the Sentinel's reasoning in observable actions. These indicators fall into two groups: pre-retrieval metrics (query analysis) and retrieval metrics (vector-space analysis). In the pre-retrieval set, \emph{Lexical Overlap} \(M_{\text{LEX}}\) measures Jaccard similarity between \(q\) and recent queries from the same user; for membership inference attacks, high overlap (e.g., \(M_{\text{LEX}} \geq 0.8\)) suggests calibrated probing behavior. \emph{Complexity Score} \(M_{\text{CMP}}\) is a ratio of non-alphanumeric characters to total characters in \(q\). \emph{Intent Velocity} \(M_{\text{INT}}\) is the inverse of the time interval between consecutive queries. In the retrieval-metric set, \emph{Vector Dispersion} \(M_{\text{DIS}}\) measures the spread of retrieved document embeddings under cosine similarity, while \emph{Score Drop-off} \(M_{\text{DRP}}\) compares the similarity scores of the top-1 and top-\(k\) retrieved documents. The Sentinel combines \(S_{\text{trust}}\), session history, and these metrics to create a \emph{risk profile}:

\begin{itemize}
    \item An overall threat level \\
    $\ell \in \{\text{LOW}, \text{ELEVATED}, \text{CRITICAL}\}$.
    \item Estimated likelihoods for each attack method, such as $p_{\text{mia}}$, $p_{\text{poi}}$, and $p_{\text{leak}}$, which relate to membership inference, data poisoning, and content leakage.
\end{itemize}

\subsubsection{Stage 2: Strategist Defense Configuration}

In the orchestration layer, the Strategist acts as a decision-making engine. It takes the Sentinel's risk profile and the current \(S_{\text{trust}}\) value and maps them to a defense plan \(\mathcal{P}\). This plan determines how the four enforcement hooks are configured. It keeps a \emph{Defense Registry} that includes differentially private retrieval (DP-RAG), TrustRAG-style clustering, and attention-variance-based leakage filters. We represent the adaptive policy as:
\[
\begin{aligned}
\pi_{\text{ado}} : &\big(q, S_{\text{trust}}, \text{history}, M_{\text{LEX}}, M_{\text{CMP}}, M_{\text{INT}}, \\
&M_{\text{DIS}}, M_{\text{DRP}}\big) \mapsto \mathcal{P},
\end{aligned}
\]
where \(\mathcal{P}\) represents, for each hook, which defenses are enabled :

\begin{itemize}
    \item \textbf{High Membership Inference Risk}: If \(p_{\text{mia}}\) is high, \(M_{\text{LEX}}\) exceeds a similarity threshold, and queries arrive at high velocity, then the Strategist configures DP-RAG with a smaller \(\epsilon\) for the upcoming retrieval stage. The retriever then injects more noise into similarity scores before top-\(k\) selection, while poisoning defenses remain at their nominal settings.
    \item \textbf{High Poisoning Risk}: The Strategist enables TrustRAG filtering at the post-retrieval hook when \(M_{\text{DIS}}\) and related dispersion indicators exceed threshold values, discarding semantic outliers before generation.
    \item \textbf{Low Risk and High Trust}: The Strategist deactivates computationally expensive defenses when the overall threat level is LOW and \(S_{\text{trust}}\) is close to 1.0.
    \end{itemize}

\subsubsection{Per-Query Orchestration Procedure}
The Sentinel-Strategist orchestration executes two LLM passes per query, as formalized in Algorithm~\ref{alg:ado}.

\textbf{Pass 1 (Pre-Retrieval).} Before retrieval, the Sentinel analyses the raw query $q$, the user's $S_\text{trust}$, and pre-retrieval metrics
$M_\text{pre}$ (lexical overlap, complexity, intent velocity) to emit an initial risk profile $\rho_\text{pre}$. The Strategist maps $\rho_\text{pre}$ to a defense plan $P_\text{pre}$ that sets any query-side transformations and configures retrieval-stage DP-RAG parameters for the upcoming retrieval call.

\textbf{Pass 2 (Post-Retrieval).} After the top-$k$ documents $D_q$ are fetched, the Sentinel refines its assessment using post-retrieval metrics
$M_\text{post}$ (vector dispersion, score drop-off) to produce $\rho_\text{post}$. The Strategist issues an updated plan $P_\text{post}$ that configures the post-retrieval hook (TrustRAG filtering), the pre-generation hook (prompt guardrails and attention-variance context pruning), and the post-generation hook (output auditing). The Global Trust Score $S_\text{trust}$ is then updated based on observed signals from the completed interaction.

\begin{algorithm}[!t]
\caption{Adaptive Defense Orchestration: Per-Query Inference}
\label{alg:ado}
\begin{algorithmic}[1]
\Require query $q$, user $u$, retrieval budget $k$
\Ensure answer $a$
\State $S_{\text{trust}} \leftarrow \textsc{TrustManager.GetScore}(u)$
\State $M_{\text{pre}} \leftarrow \textsc{ComputePreMetrics}(q, \text{history}(u))$
\State $\rho_{\text{pre}} \leftarrow \textsc{Sentinel.PreAnalyse}(q, S_{\text{trust}}, M_{\text{pre}})$
\State $\mathcal{P}_{\text{pre}} \leftarrow \textsc{Strategist}(\rho_{\text{pre}}, \texttt{stage-1})$
\State $\textsc{ActivateDefenses}(\mathcal{P}_{\text{pre}})$
\State $q', k' \leftarrow \textsc{DefenseManager.PreRetrieval}(q, k)$
\State $D_q \leftarrow \textsc{Retrieve}(\mathcal{E}(q'), \mathcal{E}(D), \text{top-}k', \mathcal{P}_{\text{pre}})$
\State $M_{\text{post}} \leftarrow \textsc{ComputePostMetrics}(D_q)$
\State $\rho_{\text{post}} \leftarrow \textsc{Sentinel.PostAnalyse}(\rho_{\text{pre}}, M_{\text{post}}, S_{\text{trust}})$
\State $\mathcal{P}_{\text{post}} \leftarrow \textsc{Strategist}(\rho_{\text{post}}, \texttt{stage-2})$
\State $\textsc{ActivateDefenses}(\mathcal{P}_{\text{post}})$
\State $D_f \leftarrow \textsc{DefenseManager.PostRetrieval}(D_q, q)$
\State $(s_{\text{sys}}, p_{\text{user}}, C) \leftarrow$
\Statex \hspace{\algorithmicindent}$\textsc{DefenseManager.PreGeneration}(q, D_f, \mathcal{P}_{\text{post}})$
\State $a \leftarrow \textsc{LLM.Generate}(p_{\text{user}}, C, s_{\text{sys}})$
\State $a \leftarrow \textsc{DefenseManager.PostGeneration}(a, \mathcal{P}_{\text{post}})$
\State $\textsc{TrustManager.Update}(u,\ \Delta = \delta_{\text{pre}} + \delta_{\text{post}})$
\State \Return $a$
\end{algorithmic}
\end{algorithm}

\section{Experiments}
\label{sec:experiments}

To evaluate our proposed Sentinel-Strategist architecture, we conducted a detailed analysis across multiple knowledge domains. We designed our experiments to measure two key outcomes: (i) the degree of semantic utility degradation caused by static, full-defense configurations, and (ii) the degree to which Adaptive Defense Orchestration (ADO) restores utility while maintaining security assurances.

\subsection{Datasets and Experimental Setup}
\label{sec:datasets}

We chose three standard benchmark datasets to evaluate our architecture, namely Natural Questions (NQ), PubMedQA, and TriviaQA. For Natural Questions (NQ), we use the dpr-w100 split from ir\_datasets to represent open-domain, real-world user queries \cite{nq,irdataset,dpr}. For PubMedQA, we adopt the pqa\_labeled configuration to model medical question answering, where accurate technical retrieval is needed \cite{pubmedqa}. For TriviaQA, we employ the rc (reading comprehension) configuration \cite{triviaqa}. Using a fixed random seed, we sample 50 benign queries from each dataset for the utility-oriented evaluation of retrieval and generation quality. These benign queries are distinct from the attack-specific poisoning and membership-inference query/probe sets described in Section~\ref{sec:threatmodel}. We execute each benign query against the 700-document ingested corpus. This sample size balances computational feasibility against the high per-query cost of the LLM-as-a-Judge evaluation pipeline; it is intentionally scoped to demonstrate the mechanistic properties of ADO rather than to characterize large-scale production performance (see Section~\ref{conclusion} for a full discussion of scope limitations).

\subsection{Implementation Details}
\label{sec:implementation}

We implement a complete RAG pipeline and the ADO framework in a modular Python architecture. As shown in Table \ref{sec:config_details}, we use \emph{Meta-Llama-3.1-8B-Instruct}\footnote{\url{https://huggingface.co/meta-llama/Llama-3.1-8B-Instruct}} as the primary generation model, with a temperature of $0.0$. Dense embeddings were generated using the \texttt{all-MiniLM-L6-v2}\footnote{\url{https://huggingface.co/sentence-transformers/all-MiniLM-L6-v2}} model \cite{minilm} and all vectors were indexed using \emph{ChromaDB}\footnote{\url{https://www.trychroma.com/}} for similarity search. We deliberately use smaller controller models for Sentinel and Strategist than one might reserve for answer generation in a production assistant, because these components consume structured telemetry and emit discrete hook configurations rather than long-form responses. In the default setup, Sentinel and Strategist use Llama-3-8B-Instruct, served via Ollama\footnote{\url{https://ollama.com/}}. We additionally evaluate the control plane with Gemma-3-4B, GPT-4o, Qwen-3-8B, and Mistral-7B to study controller sensitivity. Most local controller variants therefore remain in the compact 4B-8B range, which keeps the per-query orchestration loop more realistic for deployment by reducing incremental latency and cost even though the control plane runs on every query. All models were accessed either through Ollama or via the GPT-4o API, and all controller variants were prompted to operate over JSON-structured risk profiles and defense configurations. The Global Trust Score $S_{\text{trust}}$ is initialized to $0.5$ for all users at the start of each evaluation episode, reflecting a neutral prior, and updated per interaction via Eq.~\ref{eq:trust_update} throughout the attack sequence.

\subsection{Threat Model and Defense Configuration}
\label{sec:threatmodel}

As shown in Table \ref{attackconfigurations}, our experimental evaluation focuses on two adversarial attacks under different defense configurations: data poisoning and membership inference. For poisoning, we instantiate the \emph{PoisonedRAG} attack \cite{zou2025poisonedrag} by injecting ten adversarial documents per target query to introduce conflicting contexts, and we evaluate it on a separate set of 50 attack queries sampled with a fixed random seed. For membership inference, we instantiate the Mask-Based Attack (MBA) \cite{mba} and evaluate it on a separate 50-probe set: 30 member probes (documents present in the corpus) and 20 non-member probes (documents absent from the corpus). We use GPT-2-XL as a proxy reference model to estimate the likelihood of masked tokens without bias from the target model. Consistent with the rest of the paper, we score MBA using exact mask-fill accuracy rather than converting the reconstruction signal into a separate thresholded inference-advantage statistic.

We tailor the attack assignments to the characteristics of each dataset. We do not evaluate membership inference on Natural Questions. In PubMedQA, we focus on privacy-oriented evaluation and therefore do not run poisoning attacks. TriviaQA is used for the joint setting in which both poisoning and membership inference are evaluated. Evaluating the AV-filter path against a dedicated content-leakage benchmark, such as exact string matching or PII-style reconstruction, is left to future work; in this paper, its primary role is to complete the Sentinel-Strategist orchestration logic.

\begin{table}[h]
\caption{Core System and Retrieval Configuration}
\label{sec:config_details}
\centering
\resizebox{\columnwidth}{!}{%
\begin{tabular}{l l l}
\toprule
\textbf{Parameter} & \textbf{Value} & \textbf{Description} \\
\midrule
Generator Model & Llama-3.1-8B-Instruct & RAG generator \\
Embedding Model & all-MiniLM-L6-v2 & Dense vectorizer \\
Corpus Size & 700 & Documents per dataset \\
Chunk Size & 512 & Tokens per chunk \\
Chunk Overlap & 50 & Sliding window overlap \\
Top-$k$ & 5 & Retrieved chunks \\
\bottomrule
\end{tabular}%
}
\end{table}

\begin{table}[h]
\caption{Attack Configurations}
\label{attackconfigurations}
\centering
\resizebox{\columnwidth}{!}{%
\begin{tabular}{l l l l}
\toprule
\textbf{Attack} & \textbf{Parameter} & \textbf{Value} & \textbf{Notes} \\
\midrule
PoisonedRAG & Poisoning Rate & 10 & Docs per query \\
 & Diversity Level & True & Semantic diversity \\
Mask-based MIA & Masking Count & 5 & Tokens masked \\
 & Proxy Model & GPT-2-XL & Likelihood estimator \\
\bottomrule
\end{tabular}%
}
\end{table}

\begin{table}[h]
\caption{Defense Hyperparameters}
\label{DefenseHyperparameters}
\centering
\resizebox{\columnwidth}{!}{%
\begin{tabular}{l l l l}
\toprule
\textbf{Defense} & \textbf{Parameter} & \textbf{Value} & \textbf{Function} \\
\midrule
DP-RAG & $\varepsilon$ & 3.0 & Privacy budget \\
 & Candidate Multiplier & 3 & Oversampling \\
TrustRAG & Similarity Threshold & 0.88 & Consistency filter \\
 & ROUGE Threshold & 0.25 & Content check \\
AV Filtering & Threshold & 50 & Variance cutoff \\
 & Max Corruptions & 3 & Robustness limit \\
\bottomrule
\end{tabular}%
}
\end{table}

\subsection{Evaluation Methodology}

For our method evaluation, utility is evaluated using the DeepEval framework \cite{deepeval} with a Llama-3-base as a judge, specifically over Answer Relevancy, Faithfulness, Contextual Recall, and Contextual Relevancy. Security is measured via Attack Success Rate (ASR) for poisoning and MBA reconstruction leakage rate for membership inference. For MIA, the reported leakage rate is the exact mask-fill accuracy of the MBA attack, i.e., the percentage of evaluation probes for which the masked target content is reconstructed exactly as the reference text. Because MBA is instantiated as a reconstruction attack in our setup, we do not report a separate inference-advantage statistic; instead, we report leakage either as an aggregate rate (Table~\ref{tab:mia_efficacy}) or separately for member and non-member probe sets (Table~\ref{tab:detailed_metrics}). We used two testing protocols: a static setup (without ADO), where the benign utility queries and the attack-specific query/probe sets were evaluated sequentially, and an adaptive setup (with ADO), where inputs from those benign and adversarial pools were interleaved and randomly shuffled. Rather than relying on predictable attack patterns, this mixed evaluation ensures that the system is tested on its ability to dynamically adjust defenses per query.

\section{Results and Observations}
\label{sec:results}

\subsection{The Security-Utility Paradox}
\label{sec:utility_collapse}

We begin by measuring the \emph{utility tax} that is introduced by static defense strategies. We compare the baseline setup with no defenses against a Full Static Stack where DP-RAG, TrustRAG, and AV-Filtering are all enabled. Table~\ref{tab:utility_collapse} highlights how performance degrades across three benchmark datasets. Detailed metric-wise results and additional defense combinations are presented in Table~\ref{tab:defense-permutations}.

\begin{table}[h]
\centering
\footnotesize
\setlength{\tabcolsep}{6pt}
\caption{The Security-Utility Paradox: Impact of static defenses on retrieval performance.}
\label{tab:utility_collapse}
\begin{tabular}{lcccccc}
\toprule
& \multicolumn{2}{c}{\textbf{Context Recall}} 
& \multicolumn{2}{c}{\textbf{Faithfulness}} 
& \multicolumn{1}{c}{\textbf{Recall}} \\
\cmidrule(lr){2-3} \cmidrule(lr){4-5} \cmidrule(lr){6-6}
\textbf{Dataset} 
& Base & Full 
& Base & Full 
& \textbf{Drop} \\
\midrule
Natural Questions & 0.593 & 0.350 & 0.740 & 0.720 & \textbf{-41.0\%} \\
PubMedQA          & 0.625 & 0.335 & 0.643 & 0.720 & \textbf{-46.4\%} \\
TriviaQA          & 0.575 & 0.336 & 0.790 & 0.770 & \textbf{-41.6\%} \\
\bottomrule
\end{tabular}
\end{table}

Based on the results in Table~\ref{tab:utility_collapse}, we observe a sharp drop in retrieval utility when all static defenses are enabled. For PubMedQA, contextual recall falls by 46.4\% (0.625 $\rightarrow$ 0.335). This occurs because noise and filtering defenses disrupt high-precision retrieval. Faithfulness remains stable (within $\pm$2-3\%) across datasets, indicating that the LLM's reasoning and generation remain largely unaffected. As a result, the utility loss is driven primarily by impaired retrieval rather than by degradation of the generator itself. Table~\ref{tab:defense-permutations} further shows that configurations containing DP-RAG exhibit the largest recall drops, whereas TrustRAG-only and AV-Filtering-only remain much closer to the undefended baseline. This pattern suggests that retrieval-side perturbation is the dominant contributor to utility loss in the evaluated settings.

\subsection{Defense Effectiveness Against Targeted Threats}
\label{sec:security_efficacy}

We measure the Attack Success Rate (ASR) for data poisoning attacks and the MBA reconstruction leakage rate, defined as exact mask-fill accuracy, for membership-inference attacks.

\subsubsection{Defense Against Data Poisoning}
\label{sec:poisoning_defense}

\begin{table}[htbp]
    \centering
    \footnotesize
    \setlength{\tabcolsep}{6pt}
\caption{Poisoning mitigation across datasets using TrustRAG (CR = Contextual Recall, F = Faithfulness).}
    \label{tab:poisoning_trustrag}
    \begin{tabular}{llccc}
    \toprule
    \textbf{Dataset} & \textbf{Metric} & \textbf{No Defense} & \textbf{TrustRAG} & \textbf{Change} \\
    \midrule
    \multirow{3}{*}{TriviaQA} 
    & ASR (\%) & 56.0 & \textbf{0.0} & \textbf{-56.0\%} \\
    & CR & 0.585 & 0.574 & -1.9\% \\
    & F & 0.790 & 0.790 & 0.0\% \\
    \midrule
    \multirow{3}{*}{Natural Questions}
    & ASR (\%) & 35.0 & \textbf{0.0} & \textbf{-35.0\%} \\
    & CR & 0.598 & 0.614 & +2.7\% \\
    & F & 0.730 & 0.790 & +8.2\% \\
    \bottomrule
    \end{tabular}
    
\end{table}

As shown in Table~\ref{tab:poisoning_trustrag}, TrustRAG is highly effective at mitigating data poisoning attacks. On both TriviaQA and Natural Questions, the attack success rate (ASR) falls to zero in the evaluated setting. Relative to the full static stack in Table~\ref{tab:utility_collapse}, this targeted defense avoids the severe utility collapse caused by enabling every module simultaneously. We also observe a slight improvement in faithfulness on Natural Questions, suggesting that the defense mechanism filters out low-quality or poisoned retrieved passages.

\subsubsection{Defense Against Membership Inference}
\begin{table}[h]
\centering
\caption{Privacy Preservation: DP-RAG reduces MBA reconstruction leakage rate (exact mask-fill accuracy).}
\label{tab:mia_efficacy}
\begin{tabular}{lccc}
\toprule
\textbf{Dataset} & \textbf{No Defense} & \textbf{DP-RAG} & \textbf{Gain} \\
\midrule
TriviaQA & 29.5\% & 16.8\% & \textbf{+12.7 pp} \\
PubMedQA & 37.0\% & 19.3\% & \textbf{+17.7 pp} \\
\bottomrule
\end{tabular}
\end{table}

Based on the results in Table~\ref{tab:mia_efficacy}, we observe that DP-RAG partially mitigates Membership Inference Attacks (MIAs). Specifically, the MBA reconstruction leakage rate, measured as exact mask-fill accuracy, is reduced across datasets, dropping from 29.5\% to 16.8\% for TriviaQA and from 37.0\% to 19.3\% for PubMedQA. These findings suggest that injecting calibrated noise makes it harder for the attacker to reconstruct the masked target content exactly.

\subsection{Adaptive Restoration: The ADO Advantage}
\label{sec:ado_results}
ADO reconfigures defenses on a per-query basis through the Sentinel-Strategist pipeline described in Section~\ref{sso}. We compare it against two static baselines: a Full Static Stack that enables all defenses for every query, and a Static Targeted configuration that activates only the defense most relevant to the evaluated attack. Table~\ref{tab:ado_performance} reports aggregate results, and Table~\ref{tab:detailed_metrics} provides the per-controller, per-dataset breakdown. In the ADO rows, the model label denotes the controller used by the Sentinel-Strategist pair; the generator remains fixed. Poisoning metrics are macro-averaged over Natural Questions and TriviaQA, while MIA metrics are macro-averaged over PubMedQA and TriviaQA.

\begin{table*}[t]
\footnotesize
\centering
\caption{Comparison of ADO against static baselines on the evaluated poisoning and membership-inference settings. N/A indicates that the security metric was not evaluated.}
\label{tab:ado_performance}
\begin{tabular}{llccc}
\toprule
\textbf{Attack Scenario} & \textbf{Configuration} & \textbf{Security Metric} & \textbf{Contextual Recall} & \textbf{Faithfulness} \\
\midrule
\multirow{7}{*}{\textbf{Poisoning}}
& Full Stack (Static) & ASR: N/A & 0.343 & 0.745 \\
& Static Targeted (TrustRAG) & ASR: 0.0\% & 0.594 & 0.790 \\
& \textbf{ADO (Ours, Llama 3)} & \textbf{ASR: 0.0\%} & \textbf{0.429} & \textbf{0.753} \\
& ADO (Ours, Gemma 3) & ASR: 1.0\% & 0.297 & 0.760 \\
& ADO (Ours, GPT-4o) & ASR: 35.0\% & 0.272 & 0.770 \\
& ADO (Ours, Qwen 3) & ASR: 44.0\% & 0.439 & 0.730 \\
& \textbf{ADO (Ours, Mistral)} & \textbf{ASR: 4.0\%} & \textbf{0.470} & \textbf{0.770} \\
\midrule
\multirow{7}{*}{\textbf{MIA}}
& Full Stack (Static) & Leakage: N/A & 0.336 & 0.745 \\
& Static Targeted (DP-RAG) & Leakage: 18.05\% & 0.379 & 0.750 \\
& \textbf{ADO (Ours, Llama 3)} & \textbf{Leakage: 0.0\%} & \textbf{0.411} & \textbf{0.737} \\
& ADO (Ours, Gemma 3) & Leakage: 0.0\% & 0.282 & 0.745 \\
& ADO (Ours, GPT-4o) & Leakage: 0.0\% & 0.290 & 0.760 \\
& ADO (Ours, Qwen 3) & Leakage: 0.0\% & 0.420 & 0.720 \\
& \textbf{ADO (Ours, Mistral)} & \textbf{Leakage: 0.0\%} & \textbf{0.497} & \textbf{0.698} \\
\bottomrule
\end{tabular}
\end{table*}

\textbf{MIA Defense: Uniform Suppression of MBA Reconstruction Leakage.}
As shown in Table~\ref{tab:ado_performance}, ADO drives the aggregate MIA leakage rate to 0.0\% across all five evaluated controller variants. In the per-dataset breakdown of Table~\ref{tab:detailed_metrics}, all five variants also achieve 0.0\% leakage on both member and non-member probe sets. This improves on the static targeted DP-RAG baseline, which still leaves 18.05\% aggregate leakage, while restoring contextual recall to 0.411 with the Llama 3 controller and 0.497 with the Mistral~(7B) controller.

\textbf{Poisoning Defense and Inter-Model Variance.} Poisoning results are more controller-sensitive. The Llama 3 and Mistral controllers keep ASR at 0.0\% and 4.0\%, respectively, while recovering contextual recall to 0.429 and 0.470, approaching the static targeted TrustRAG upper bound of 0.594. GPT-4o and Qwen~3 are less reliable, reaching 35.0\%-44.0\% aggregate ASR and up to 50.0\% on Natural Questions, suggesting looser activation of TrustRAG under borderline cases. Gemma~3 largely suppresses poisoning while restoring the least utility. Overall, ADO shows strong MBA suppression in the evaluated membership-inference setting, whereas poisoning robustness remains controller-dependent. Content leakage remains architecturally represented through the AV-filter path rather than benchmarked separately.

\begin{table*}[t]
\centering
\caption{Utility scores for all defense permutations across datasets (CR = Contextual Recall, CP = Contextual Relevancy, AR = Answer Relevancy, F = Faithfulness).}
\label{tab:defense-permutations}

\resizebox{\textwidth}{!}{%
\begin{tabular}{@{}l cccccccccccc@{}}
\toprule
 & \multicolumn{4}{c}{\textbf{Natural Questions}} & \multicolumn{4}{c}{\textbf{PubMedQA}} & \multicolumn{4}{c}{\textbf{TriviaQA}} \\
\cmidrule(lr){2-5} \cmidrule(lr){6-9} \cmidrule(lr){10-13}
 & CR & CP & AR & F & CR & CP & AR & F & CR & CP & AR & F \\
\midrule
Baseline (none) & 0.593 & 0.489 & 0.930 & 0.740 & 0.625 & 0.424 & 0.833 & 0.643 & 0.575 & 0.514 & 0.700 & 0.790 \\
DP-RAG only (D) & 0.361 & 0.321 & 0.863 & 0.770 & 0.398 & 0.270 & 0.793 & 0.680 & 0.359 & 0.264 & 0.750 & 0.820 \\
TrustRAG only (T) & 0.586 & 0.470 & 0.860 & 0.720 & 0.584 & 0.418 & 0.803 & 0.712 & 0.567 & 0.509 & 0.720 & 0.810 \\
AV-Filtering only (A) & 0.592 & 0.449 & 0.780 & 0.770 & 0.591 & 0.390 & 0.760 & 0.666 & 0.606 & 0.513 & 0.680 & 0.830 \\
D + T & 0.380 & 0.344 & 0.853 & 0.670 & 0.383 & 0.267 & 0.803 & 0.660 & 0.339 & 0.266 & 0.790 & 0.750 \\
D + A & 0.364 & 0.286 & 0.740 & 0.580 & 0.354 & 0.224 & 0.563 & 0.700 & 0.354 & 0.222 & 0.740 & 0.760 \\
T + A & 0.571 & 0.387 & 0.750 & 0.690 & 0.520 & 0.326 & 0.477 & 0.770 & 0.590 & 0.494 & 0.713 & 0.790 \\
\textbf{Full Stack (D+T+A)} & \textbf{0.350} & \textbf{0.305} & \textbf{0.863} & \textbf{0.720} & \textbf{0.335} & \textbf{0.205} & \textbf{0.583} & \textbf{0.720} & \textbf{0.336} & \textbf{0.232} & \textbf{0.690} & \textbf{0.770} \\
\bottomrule
\end{tabular}%
}
\end{table*}

\begin{table*}[t]
\centering
\footnotesize
\begin{threeparttable}
\caption{Per-controller-model utility and security results. CR = Contextual Recall, CP = Contextual Relevancy, AR = Answer Relevancy, F = Faithfulness; ASR = poisoning attack success rate; Leak. = membership-inference leakage; N/A = not evaluated.}
\label{tab:detailed_metrics}
\begin{tabular}{llccccccc}
\toprule
\textbf{Controller} & \textbf{Dataset} & \textbf{AR} & \textbf{F} & \textbf{CR} & \textbf{CP} & \textbf{ASR} & \textbf{Leak. (Mem.)} & \textbf{Leak. (Non.)} \\
\midrule
\multirow{3}{*}{\textbf{Llama 3 (8B)}} & PubMedQA & 0.730 & 0.723 & 0.347 & 0.587 & N/A & 0.0\% & 0.0\% \\
 & NaturalQ & 0.897 & 0.755 & 0.382 & 0.523 & 0.0\% & N/A & N/A \\
 & TriviaQA & 0.810 & 0.750 & 0.475 & 0.539 & 0.0\% & 0.0\% & 0.0\% \\
\midrule
\multirow{3}{*}{\textbf{Gemma 3 (4B)}} & PubMedQA & 0.723 & 0.720 & 0.241 & 0.376 & N/A & 0.0\% & 0.0\% \\
 & NaturalQ & 0.770 & 0.750 & 0.271 & 0.365 & 2.0\% & N/A & N/A \\
 & TriviaQA & 0.770 & 0.770 & 0.322 & 0.345 & 0.0\% & 0.0\% & 0.0\% \\
\midrule
\multirow{3}{*}{\textbf{GPT-4o}} & PubMedQA & 0.763 & 0.760 & 0.290 & 0.393 & N/A & 0.0\% & 0.0\% \\
 & NaturalQ & 0.810 & 0.780 & 0.255 & 0.341 & 36.0\% & N/A & N/A \\
 & TriviaQA & 0.750 & 0.760 & 0.289 & 0.326 & 34.0\% & 0.0\% & 0.0\% \\
\midrule
\multirow{3}{*}{\textbf{Qwen 3 (8B)}} & PubMedQA & 0.850 & 0.750 & 0.364 & 0.469 & N/A & 0.0\% & 0.0\% \\
 & NaturalQ & 0.883 & 0.770 & 0.402 & 0.513 & 50.0\% & N/A & N/A \\
 & TriviaQA & 0.830 & 0.690 & 0.475 & 0.493 & 38.0\% & 0.0\% & 0.0\% \\
 \midrule
\multirow{3}{*}{\textbf{Mistral (7B)}} & PubMedQA & 0.830 & 0.685 & 0.482 & 0.600 & N/A & 0.0\% & 0.0\% \\
 & NaturalQ & 0.930 & 0.830 & 0.427 & 0.564 & 4.0\% & N/A & N/A \\
 & TriviaQA & 0.850 & 0.710 & 0.512 & 0.555 & 4.0\% & 0.0\% & 0.0\% \\
\bottomrule
\end{tabular}
\vspace{0.25em}
\parbox{\linewidth}{\footnotesize $^\dagger$ 0.0\% leakage under ADO indicates that MBA did not exactly reconstruct the masked target span on the evaluated probes.}

\end{threeparttable}
\end{table*}

\balance
\section{Conclusion and Future Work}
\label{conclusion}

This work exposes a security-utility paradox in RAG systems: an always-on defense stack sharply reduces contextual recall by 41-46\% even when faithfulness remains stable, indicating that the main failure mode is impaired retrieval rather than degraded generation. We address this with Adaptive Defense Orchestration (ADO), in which a Sentinel forms per-query risk profiles and a separate Strategist maps them to hook-level defenses. This split keeps detection and actuation auditable and allows the control plane to run on compact controller models. Across five controller-model variants and three benchmark datasets with attack-specific pairings, ADO suppresses MBA reconstruction leakage in the evaluated membership-inference setting and recovers retrieval utility relative to the static full stack, although poisoning robustness remains model-sensitive. Content leakage is represented through the AV-filter path but is not benchmarked separately here.

\paragraph*{Limitations and Future Work}
Our evaluation is intentionally scoped to feasibility: 50 benign queries and 50 adversarial queries per attack class per dataset, a 700-document corpus, single-run point estimates, no dedicated content-leakage benchmark, and no systematic temporal or multi-turn attack patterns. The Sentinel-Strategist control plane introduces additional orchestration overhead per query. We do not report a fixed end-to-end latency multiplier because the observed cost varies with the inference backend, model serving configuration, and the defense intensity selected for a given query. In practice, ADO activates defenses on most queries, but calibrates their aggressiveness to the assessed threat level rather than applying the full static stack uniformly. Next steps include temporal and multi-turn attack studies, model-specific prompt calibration, Sentinel distillation, and more efficient parallel defense activation.

\bibliographystyle{unsrtnat}
\bibliography{references}

\appendix

\section{Open Science}
The core contributions of this paper depend on the following artifacts: (i) the RAG pipeline implementation, (ii) Sentinel and Strategist prompts together with the control registry, (iii) attack scripts for PoisonedRAG and the Mask-Based Attack (MBA), (iv) evaluation scripts, configuration files, and random seeds, and (v) instructions for reproducing the reported tables and figures. For double-blind review, these artifacts are provided through the following anonymous archive:

\begin{quote}
\url{https://github.com/Pranavdec/Adaptive-RAG-Orchestrator}
\end{quote}

The anonymous archive maps the included files to the paper's experiments and tables, documents the required software dependencies, and explains any steps needed to reproduce the reported results. Third-party model weights, external APIs, and access credentials are not redistributed in the archive; instead, we document the exact model identifiers, configuration settings, and interfaces used in the experiments.

\section{Ethical Considerations}
This paper studies membership-inference and data-poisoning attacks against retrieval-augmented generation systems in order to evaluate defensive orchestration strategies. Because the work discusses attack mechanisms, releasing artifacts may lower the barrier to misuse. We therefore evaluate all attacks in a controlled offline setting over benchmark datasets and locally constructed corpora, and we do not test against third-party production systems or private user deployments.

Our experiments do not involve human-subject interaction, and the reported datasets are public benchmarks commonly used for question answering research. The study does, however, concern privacy and integrity risks for systems that may store sensitive documents. To reduce the chance of harm, we frame the attacks only as evaluation instruments for measuring defense effectiveness, avoid including live targets or deployment-specific secrets, and separate reusable configuration details from any sensitive credentials or infrastructure information. The goal of releasing the methodology is to support reproducible defensive research rather than to operationalize attacks against real users.

\end{document}